\documentstyle[12pt]{article}
\def\be{\begin{equation}}
\def\ee{\end{equation}}
\begin{document}
\title{\bf{Phase Transition in Warm Nuclear Matter with Alternative 
 Derivative Coupling Models}}  
\vskip .3cm
\author{ M. Malheiro $^{1,2}$ \thanks{Partially
supported by CAPES of Brasil},
 A. Delfino $^{1}$ \thanks{Partially supported by CNPq of 
Brasil} 
 and C. T. Coelho$^{1}$\\ \\ $^{1}$ Instituto de 
F\'\i sica, Universidade Federal Fluminense, \\ 24210-340, Niter\'oi,
R. J., Brasil
\\
${^2}$ Department of 
Physics, University of Maryland, \\ College Park, Maryland 20742-4111, USA}
\vspace{.3 cm} 
\date{\today}
\maketitle
{\it Abstract} :
An analysis is performed of the liquid-gas phase transition of nuclear
matter obtained from different versions of scalar derivate coupling 
suggested by Zimanyi and Moszkowski (ZM) and the results are
compared with those obtained from the Walecka model. We present the phase 
diagram for the models and one of them,  
the ZM3 model, has the lowest critical temperature
$T_c=13.6$ MeV with the lowest critical
density $\rho_c=0.037$ f$m^{-3}$ and pressure $p_c=0.157$ MeV f$m^{-3}$.
 These results are in accord with recent observations from energetic heavy-ion 
collisions, which suggest a small liquid-gas phase region.
\\
PACS  \# 21.65.+f, 12.40.Yx, 25.75.+r

\newpage  
\section {  Introduction}
\hskip 0.7cm
   
   Nowadays, the study of the liquid-gas phase transition, which may occur
in the warm and dilute matter produced in energetic heavy-ion collisions,
is one of the interesting problems in nuclear physics \cite{li}. This idea of that 
nuclear systems may show up a critical behavior has initiated more than ten years ago with the observation by the Purdue-Fermilab group of asymptotic fragment charge distributions exhibiting a power law \cite{pur}. This interest
increased recently with the attempt by the EOS Collaboration to extract critical
 exponents of fragmenting nuclear systems produced in the collision of 1 GeV/nucleon Au nuclei with a carbon target \cite{pur1}, and with the extraction by ALADIN/LAND Collaboration of a caloric curve resulting from the fragmentation
 of the quasiprojectile formed in the collision Au + Au at 600 MeV/nucleon exhibiting a behavior expected for a  first order liquid-gas phase transition \cite{jp}.

 At the time where the search for signals of liquid-gas nuclear phase transition  are taking place, it is important to have ready the theoretical phase-transition predictions for a broad class of different hadronic models. The main ingredient in this analysis is the nuclear matter equation of state (EOS) at finite temperature.
The success of relativistic mean-field theories describing cold
nuclear matter and bulk nuclear properties throughout the periodic table,
suggests the use of a relativistic  mean-field EOS. Moreover, 
the mean-field approximation is known to be thermodinamicaly consistent: 
the relevant thermodynamic identities are all satisfied \cite{furs}, \cite{mb}.
   
   Recently variants of Zimanyi and Moszkowski (ZM) model \cite{zm}
 were implemented  and applied by us to dense and cold nuclear matter
 \cite{acm1}, \cite{acm2}.
 The usual ZM model, also  referred to in the literature 
as the Derivative Scalar Coupling model (DSC), consists of  derivative
coupling between nucleons and scalar mesons $\sigma$. The model
has been extended to include a non-linear
interaction between the nucleon and the vector meson $\omega$. Two types
of this interaction were employed and the resulting models were denoted
ZM2 and ZM3. 
These models were designed to cure the defects of the Walecka model \cite{sw},
namely the low effective nucleon mass and the large incompressibility
of nuclear matter.
Each one of them is very
simple since they have only two free parameters, the scalar (vector)
coupling constants $C_{\sigma}^2$ $(C_{\omega}^2)$, adjusted to reproduce the 
binding energy ($E_b$) of the nuclear matter at $\rho = \rho_0$. 
The degrees of freedom  are baryon fields ($\psi$), scalar meson fields
 ($\sigma$), and vector meson fields ($\omega$).

In all ZM models, there are non-linear interaction terms which in a approximate
way incorporate the effect of many-body forces. After an
appropriate rescaling of the Lagrangians, these models can be
understood as generalizations of the Walecka Model where the scalar and vector
meson couplings become effectively density-dependent \cite{acmbg}. This fact
 underlies the recent approach, known as relativistic 
density-dependent Hartree-Fock \cite{scm}, \cite{lz}, \cite{sg}  which
 describes finite nuclei and nuclear matter saturation properties
 using coupling constants that are fitted, at each density value, to the
 relativistic Brueckner-Hartree-Fock self-energy terms. The good agreement 
obtained for
the ground state properties of spherical nuclei lends support to such
 density-dependent coupling constants. Recently, it was shown
that chiral symmetry restoration requires the meson-nucleon coupling
to be density dependent \cite{addm}.

The aim of this paper is to extend our study to include  
temperature effects, and to perform
an analysis of the liquid-gas phase transition of the warm nuclear
matter obtained on these three ZM models and compare to  
the linear Walecka (W) model \cite{inpc}. We 
 present the effective nucleon mass, energy per nucleon, 
pressure, and entropy  density  as a function of the baryonic density
 at different
 temperatures. We show the isotherms, construct the phase diagram with 
 the phase coexistence
 boundary, and present the critical  
 and flash temperatures for the models.
 The usual ZM model has already been applied to 
investigate some thermodynamic properties of nuclear matter \cite{qss} and
 recently, the modified versions have been used in order to study 
 the density and temperature dependence of hadron masses \cite{br}. 

The outline of the paper is as follows: in the next section we
present the EOS at finite temperature. Section 3 includes
our results and discussion of the thermodynamic properties of nuclear matter.
Finally, we summarize.

\section {  The nuclear matter EOS at finite temperature}
%\hskip 0.5cm

Since the models we are dealing with were discussed in detail in
references \cite{zm}, \cite{acm1}, \cite{acm2}, here we will only present 
 the Lagrangian obtained after rescaling the nucleon field as $\psi
\rightarrow {m^{\ast}}^{1/2}\psi$  for all ZM models and making the rescaling  
 $\omega_{\mu} \rightarrow m^{\ast}\omega_{\mu}$  for  ZM2 and ZM3 models:
\begin{eqnarray}
\label{lu}
{\cal L}_{R} & = & \bar \psi i \gamma_{\mu} \partial^{\mu } \psi  
 +  \ m^{\ast^{\alpha}}\ \left(  - g_{\omega}\bar \psi \gamma_{\mu }\psi
\omega^{\mu } - \frac{1}{4}F^{\mu \nu}F_{\mu \nu} + \frac{1}{2} 
m^2_{\omega}\omega_{\mu}\omega^{\mu}\right) \nonumber\\
& - & \bar \psi( M - m^{\ast^{\beta}} g_{\sigma}\sigma ) \psi   
+ \frac{1}{2}(\partial_{\mu }\sigma\partial^{\mu}\sigma - m^2_{\sigma}
\sigma^2), 
\end{eqnarray}
where $\alpha$ and $\beta$ have the following values
for the different models,
$ W: \alpha = 0 , \beta = 0 $; $ ZM: \alpha = 0 , \beta = 1 $;
 $ ZM2: \alpha = 1 , \beta = 1 $; $ ZM3: \alpha = 2 , \beta = 1 $;
 and $  m^{\ast} = (1 + g_{\sigma}\sigma/M)^{-1} $  in all three cases, 
 M is the bare nucleon mass
and $F_{\mu \nu} = \partial_{\mu}\omega_{\nu} - \partial_{\nu}\omega_{\mu} 
$.

When the meson fields are replaced by the constant classical fields
$\sigma_{o}$ and $\omega_{o}$ we arrive at the mean-field
approximation, with the equation of motion for the nucleon:
\begin{equation}
\label{cnu}
\ [i \gamma_{\mu} \partial ^{\mu} - ( M -
m^{\ast^{\beta}} g_{\sigma} {\sigma}) - m^{\ast^{\alpha}} g_{\omega}
\gamma_{\mu} \omega^{\mu} ] \psi = 0, 
\end{equation} 
where the effective nucleon mass $M^{\ast}$ is given by  $M^* = M - m^{*\beta} 
g_\sigma \sigma $. In the case
of ZM models where $ \beta = 1 $ we can identify 
$ m^{\ast} =  M^{\ast}/M = (1 + g_{\sigma}\sigma/M)^{-1} $.

The expression for the energy density and pressure at a given
temperature T can be found as usual by the average of the
energy-momentum tensor,
\begin{equation}
\label{enert}
{\cal E} = \frac{C_{\omega}^{2}}{2M^{2}}m^{\ast^{\alpha}}\rho^2+
\frac{M^{4}}{2C_{\sigma}^{2}}\left(\frac{1-m^{\ast}}{m^{\ast{\beta}}}\right)^2+
\frac{\gamma}{(2\pi)^{3}}\int d^{3}k\,E^{\ast}(k)(n_{k}+\bar n_{k}) ,
\end{equation}
\begin{equation}
\label{prest}
 p = \frac{C_{\omega}^{2}}{2M^{2}}m^{\ast^{\alpha}}\rho^2 -
\frac{M^{4}}{2C_{\sigma}^{2}}\left( \frac{1-m^{\ast}}{m^{\ast^{\beta}}}\right)^2 +
\frac{1}{3}\frac{\gamma}{(2\pi)^{3}}\int d^{3}k\,\frac{k^{2}}{E^{\ast}(k)}
(n_{k}+\bar n_{k}) .
\end{equation} 
Thus we obtain the entropy density:
\begin{eqnarray}
\label{entrt}
s & = & \frac{1}{T} \left[ \frac{ C_\omega^2}{M^2} m^{*\alpha} \rho^2 + \frac{\gamma}
{(2\pi)^{3}}\int d^{3}k\,E^{\ast}(k)(n_k+\bar n_{k}) \right]
\nonumber\\ \nonumber\\
& + &\frac{1}{3T} \frac{\gamma}{(2\pi)^{3}}\int d^{3}k\,\frac{k^{2}}{E^{\ast}(k)}
(n_{k}+\bar n_{k}) - \frac{\mu \rho}{T},
\end{eqnarray}
where $ \gamma$ is the degeneracy factor ( $\gamma = 4 $ for nuclear
matter and $ \gamma = 2 $ for pure neutron matter ), $n_{k}$ and $\bar n_{k}$
stand for the Fermi-Dirac distribution for baryons and antibaryons 
respectively,
with arguments $ (E^{\ast} - \nu)/T $ , $E^{\ast}(k)$ is given by
$ E^{\ast}(k) = ( k^2 + M^{\ast 2} )^{\frac{1}{2}} $ . An effective chemical
potential which preserves the number of baryons and antibaryons in the 
ensemble is defined by $ \nu=\mu - V $, with $\mu$ is the thermodynamical
chemical potential. We have introduced
 $C_{\sigma}^2 = g_{\sigma}^2M^2/m_{\sigma}^2$  and
 $C_{\omega}^2 = g_{\omega}^2M^2/m_{\omega}^2$.
 
 The effective mass is obtained explicitly
through the minimization of ${\cal E }$ with respect to $m^{\ast}$ and must
satisfied the self-consistent equation,
\begin{equation}
\label{meft}
1 - m^{\ast} - \frac{\gamma C_\sigma^2}{2\pi^2} m^{\ast^{3\beta+1}}\int\,
\frac{x^2dx}{\sqrt{x^2+m^{\ast^2}}}(n_{x}+\bar n_{x})-\frac{\alpha}{2}
\frac{C_{\sigma}^2C_{\omega}^2}{M^6}m^{\ast^{\alpha+2\beta}}\rho^2=0
\end{equation}
where we have used the dimensionless variable $ x=\frac{k}{M} $. 

The energy density can be fitted to the nuclear matter ground state
energy and saturation density $\rho_{o}$ at zero temperature 
to obtain the different coupling constants for the models. 
They are presented in table 1 together with the nuclear matter 
 incompressibility that at $T=0$ is given by:
\begin{equation}
\label{k}
K=9\rho_{o}^2\frac{\partial^2}{\partial\rho^2}(\frac{{\cal E}}{\rho})
\vert_{\rho = \rho_o}=9\rho_{o}\frac{\partial^2{\cal E}}{\partial\rho^2}
\vert_{\rho = \rho_o}.
\end{equation}

To compute the thermodynamic functions, one first chooses T and $ \nu$.
 The self-consistency condition in eq.(\ref{meft}) is then solved to determine
 $M^*$ ( note that there may be several solutions for fixed T and $ \nu$ ).
These solutions specify the distribution functions $n_{k}$ and $\bar n_{k}$,
 and the remaining integrals in Eqs. (\ref{enert}), (\ref{prest}) and 
 (\ref{entrt}) can then be evaluated directly.

\section{ Results and Discussion}      
\hskip 0.7cm

In Fig. 1 we show $M^{\ast}$ as a function 
of T at zero density. 
In this regime, the vector field proportional to $\rho$
 vanishes, and so the three ZM models differ only in having different values of
 their scalar coupling constants $C_{\sigma}^2 $. The ZM 
 and the Walecka model coincide in the lower temperature region $T \le
120$ MeV and  ZM3 model stay together up to $T \sim 160$ MeV. However,
at higher temperature the models separate quite clearly, with the effective 
nucleon mass in the ZM models dropping more slowly than that in the 
Walecka Model. This means that the sigma field (the  
source for the scalar density), increases more slowly with temperature
in ZM models because of the inclusion of non linear interactions which are
absent in the Walecka Model. As a result, the attraction is stronger in 
 the Walecka model favouring the formation of nucleon-antinucleon pairs at
high temperature.
 Moreover,  none of
the proposed ZM models is able to present a first order phase transition at
 $\rho$= 0 , $T\ne$0. This is in contrast to the Walecka model, which has such
 a phase transition at  $T \sim 185$ MeV \cite{tsp}.
 
In Fig. 2 we show the behavior of the effective nucleon mass with
density at different temperatures for all the models. For low temperatures
the results are not so different from those obtained at zero temperature,
 showing that in this regime the density dependence is more important
than the temperature dependence. As temperature is raised,
$M^*$ first increases  and then decreases more slowly for ZM
models than Walecka model at $T=200$ MeV. Within ZM models, this decrease
is more pronounced in ZM3, but is even smaller compared to Walecka model
 where the effective mass goes down very fast.
In short, the effect of the
temperature on the effective nucleon mass in the ZM models
is not so pronounced as in the case of Walecka Model, and can be seen
only for densities below the normal density. 

We present the energy per nucleon as a function of the density at
various temperatures in Fig. 3 . As
the temperature increases the nuclear matter becomes less bound and the
the saturation curve around the equilibrium point in the
ZM models is flatter than that in the Walecka model. This indicates
that the nuclear matter EOS in ZM models is softer compared 
to the obtained in the Walecka Model, even at finite temperature. 
We can also conclude that the incompressibility of nuclear matter decreases 
when the temperature increases. This can be seen more clearly in Fig. 4 
where we show the pressure-density isotherms of nuclear matter at
different temperatures. Since the incompressibility K is related
 to ${\partial p}/{\partial\rho}$ (calculated at the
equilibrium point where the pressure vanishes), we see directly that
when the temperature increases K decreases, and  among the ZM models,
the ZM3 model always gives the softest EOS for a fixed temperature.

The isotherms exhibit  a typical Van der Waals like interaction
where a liquid and gaseous phases coexist, with an unphysical
region in the midle of each isotherm that gets smaller as the
temperature increases. For very small temperatures the isotherms manifest 
the following behavior: for very low density the pressure increases
with temperature as for as a ideal gas, $ p \sim \rho k_b
T$. It decreases subsequently because
of the attractive interaction of the sigma field, and finally increases
 as a consequence of the repulsion coming from the vector meson
which dominates at high density.
When temperature increases, the term $ \rho k_bT$ becomes
more important and the local minimum in the pressure is less pronounced
and disappears when the temperature is equal to the critical 
$T_c$. At this temperature, the unphysical region disappears and an inflection
 point appears in the isotherm, as we show in the Fig.4 for each model. 
The p-$\rho$ isotherms in the ZM models have a shallower and more
even valley than the corresponding ones in the Walecka Model, and this
is more noticeable in  ZM3 model.
 In table 2 we list the critical temperature $T_c$,
density $\rho_c$ and pressure $p_c$ given by the ZM  and Walecka models.
The ZM3 model presents the lowest $T_c=13.6$ MeV,
density $\rho_c=0.037$ f$m^{-3}$ and pressure $p_c=0.157$ MeV f$m^{-3}$.

The phase coexistence boundary is determined by Gibbs's criteria, namely,
 that the liquid and gas phases have equal temperatures (thermal
equilibrium), chemical potentials (chemical equilibrium), and
pressures (hydrostatic equilibrium). We present in Fig.5 the phase diagram
Tx$\rho$ of the models. Below the coexistence curve of each one, the
equilibrium state is a mixture of gas and liquid. This region is
bigger in the Walecka model. In fact, if we include nonlinear terms
in this model, this region becames smaller and the critical temperature 
goes down to $T_c=14.2$ MeV \cite{furs}. The ZM3 model, where the 
non-linearity of the coupling
 between the vector field to the nucleon is strongest, 
presents the smallest
phase coexistence region comparing to the other models.
 
As we have already pointed out, the nuclear matter incompressibility
K decreases when the temperature increases. So, we will have a temperature
where the incompressibility K calculated at the equilibrium point 
vanishes.
 This temperature is known as the flash temperature $T=T_{f}$, 
 $\frac{\partial p}{\partial\rho}\vert_{
T_f}\,=\,p(\rho_f,T_f)=0 $. It represents the highest temperature
at which a self-bound system can exist in hydrostatic equilibrium ($p=0$).
 Above this temperature the warm nuclear matter is unbound 
and starts expanding.
We present in Fig. 6  the pressure as a function of baryon density at
the flash temperature for the models. This temperature in MeV is 14.1, 12.9,
 12.2 and 11.0 for the Walecka, ZM, ZM2 and ZM3 models respectively.
 Again, the ZM3 model has the smallest flash temperature.
 As expected, all of these temperatures are lower than the critical ones,
   because, as Fig.4 shows, at the critical temperature
 the pressure is already positive and the system is expanding.
 
Finally, we present in Fig.7 the entropy density as a function of the density
 at different temperatures. For high temperatures ($T=200$ MeV), we
see an increase in the entropy density with the density for all the models. This
happens even at very low densities, and manifests what we have already
pointed out when we have discussed the behavior of the effective
nucleon mass with the temperature at zero density. This 
decrease of $M^*$ or increase of the entropy density with increasing
temperature, which is more pronounced in the Walecka and ZM3 models,
resembles a phase transition. At high temperature and
low density the system becomes a dilute gas of baryons in a sea of 
baryons-antibaryons. 

In summary,
 we have presented the thermodynamic properties of nuclear matter in three
different versions of the ZM model. We have shown how the effective nucleon
 mass $M^{\ast}$, energy per nucleon, pressure,
 and entropy 
behaves as a function of the density for different temperatures. 
As in the zero temperature case, all the ZM models
give a softer EOS of nuclear matter at finite temperature than the
Walecka model. Among the three ZM models, ZM3 is the softest.
Unlike the Walecka Model the ZM models do not exhibit a phase
transition for finite temperature at zero density.
We studied the liquid-gas phase transition, and concluded
that the ZM3 model presents the smallest
phase coexistence region with
the lowest critical temperature, density and pressure. The incompressiblity
decreases with the increasing temperature,
and vanishes when $T=T_{flash}$. Again, the ZM3 model has the smallest
flash temperature. The experimental results investigating this warm and dilute 
matter produced 
in energetic heavy-ion collisions suggest a small liquid-gas 
phase region and a low critical temperature \cite{jp}, \cite{li}. This 
suggests that the good 
description of nuclear matter properties  obtained in the ZM3 model at 
zero temperature remains, even at finite temperature, and makes the ZM3 
model the most suitable of all models used.

One of us (MM) would like to thank Dr. D. R. Phillips for careful reading of the manuscript, the Maryland TQHN Group supported by the U.S. Depto. of Energy for their hospitality during his extended stay, and gratefully the CAPES of the Brazilian Governement for the financial support which made this visit possible.

\newpage
\section{Figures and Captions }
Fig. 1: Baryon effective mass in nuclear matter as a function of the
temperature at $\rho=0$.
\\
\\
\\
Fig. 2: Baryon effective mass $M^*$ as a function of the 
baryon density at different temperatures for the Walecka model (W)
and Zimanyi-Moszkowski models (ZM, ZM2, ZM3).
\\
\\
\\
Fig. 3: Proper energy/baryon as a function of baryon density at different
temperatures for the Walecka model (W)
and Zimanyi-Moszkowski models (ZM, ZM2, ZM3).
\\
\\
\\
Fig. 4: Pressure as a function of baryon density at different
temperatures for the Walecka model (W)
and Zimanyi-Moszkowski models (ZM, ZM2, ZM3).
\\
\\
\\
Fig. 5: Temperature as a function of the baryon
density (phase diagram) for the Walecka model (W)
and Zimanyi-Moszkowski models (ZM, ZM2, ZM3).
\\
\\
\\
Fig. 6: Pressure as a function of baryon density at
flash temperature ($T_f$) for the Walecka model (W) and
Zimanyi-Moszkowski models (ZM, ZM2, ZM3).
\\
\\
\\
Fig. 7: Entropy density as a function of baryon density at different
temperatures for  the Walecka model (W)
and Zimanyi-Moszkowski models (ZM, ZM2, ZM3).

\newpage
\begin{table}
\caption{Coupling constants $ C_\sigma^2 $ and $ C_\omega^2 $;
binding energy $ E_b $
(MeV) at equilibrium density $\rho_o (fm^{-3}) $,
$ m^{\ast} $ and the incompressibility K for the indicated models.}
\label{models}
\begin{center}
\begin{tabular}[pos]{|c|c|c|c|c|c|c|} \hline
\,\,models \,\, &\,\,\,\,\, $C_\sigma^2$ \,\,\,\,\, & \,\,\
\,\, $C_w^2$ \,\,\,\,\,&  \,\,\,\,\,\, $E_b$\,\,\,\,\,\, &\,\,\,\,\,\,\
$\rho_o$\,\,\,\,\,\, &\,\,\,\,\,\,$m^*$\,\,\,\,\,\,&\,\,\,\,\ $ K $\,\,\,\,  \\ \hline
W   & 357.4      & 273.8 & -15.75 & 0.148 & 0.54 & 550.82\\ \hline 
ZM  & 169.2 & \'59.1 & -15.90 & 0.160 &0.85 & 224.71\\ \hline
ZM2 & 219.3 & 100.5 & -15.77 & 0.152 & 0.82 & 198.32\\ \hline
ZM3 & 443.3 & 305.5 & -15.76 & 0.149  &0.72 & 155.74 \\ \hline 
\end{tabular}
\end{center}
\end{table}

\begin{table} 
\caption{Values for the critical temperature $T_c$ and the effective mass
 $M^*_c$ in MeV, critical density $\rho_c$ in $fm^{-3}$ and pressure $p_c$
in MeV/ $fm^{3}$ for the indicated models.}

\label{tc}
\begin{center}
\begin{tabular}[pos]{|c|c|c|c|c|} \hline
\,Models\, & \,\, $T_c$ \,\, & \,\,\,\ $\rho_c$\,\,\,\,\, & \,\, $p_c$\,\, &
\,\, $M_c^*$ \,\,  \\ \hline
 Walecka & 18.3  & 0.0650  & 0.4300 & 760 \\ \hline
 ZM  & 16.5 & 0.0698 & 0.2570 & 861 \\ \hline
 ZM2 & 15.5 & 0.0364 & 0.2106 & 881  \\ \hline
 ZM3 & 13.6 & 0.0354 & 0.1571 & 831  \\ \hline
\end{tabular}
\end{center}                     
\end{table}
  
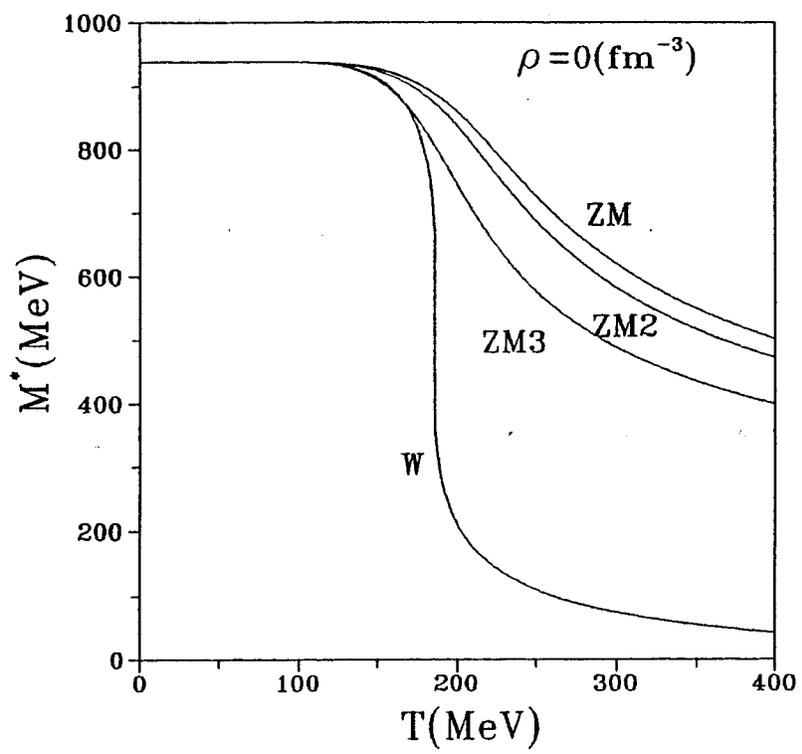
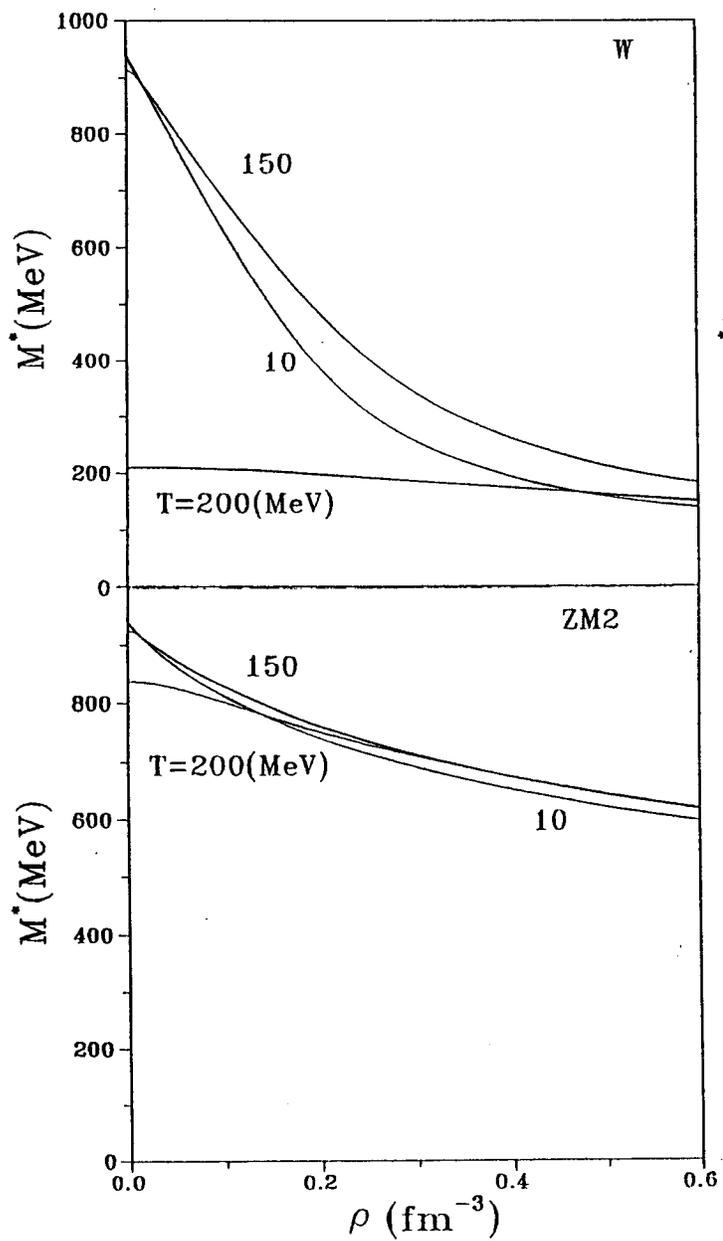
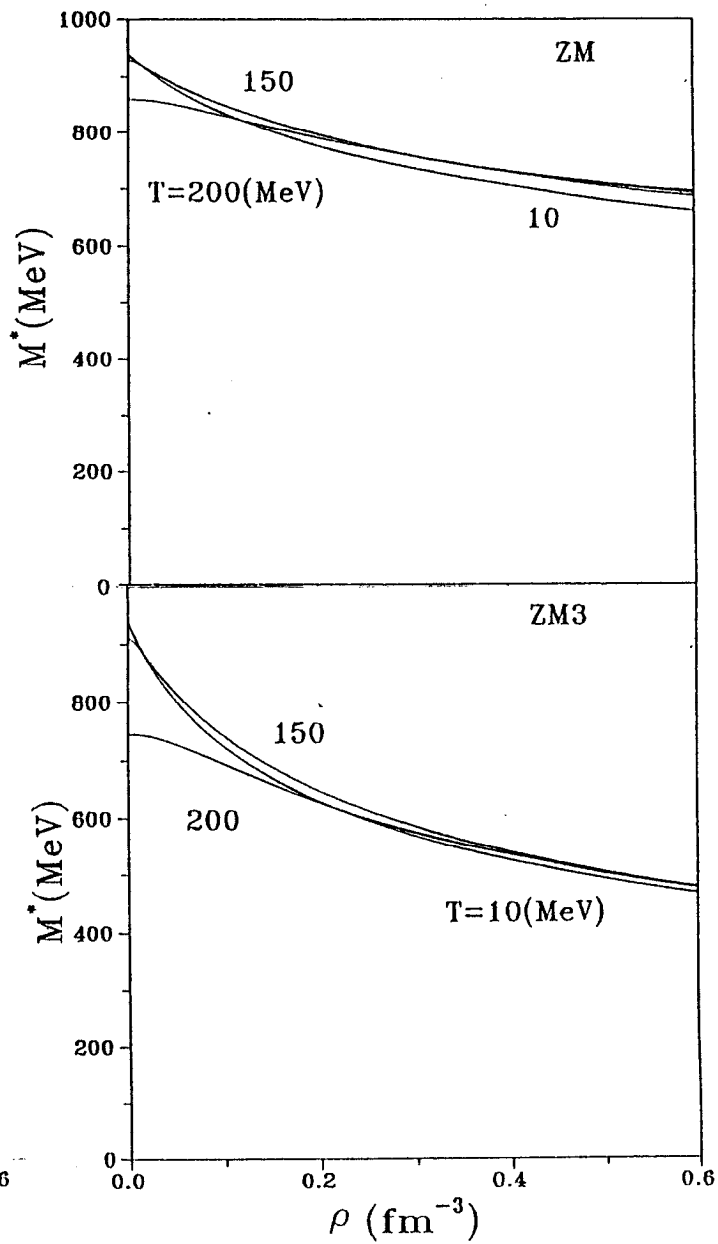
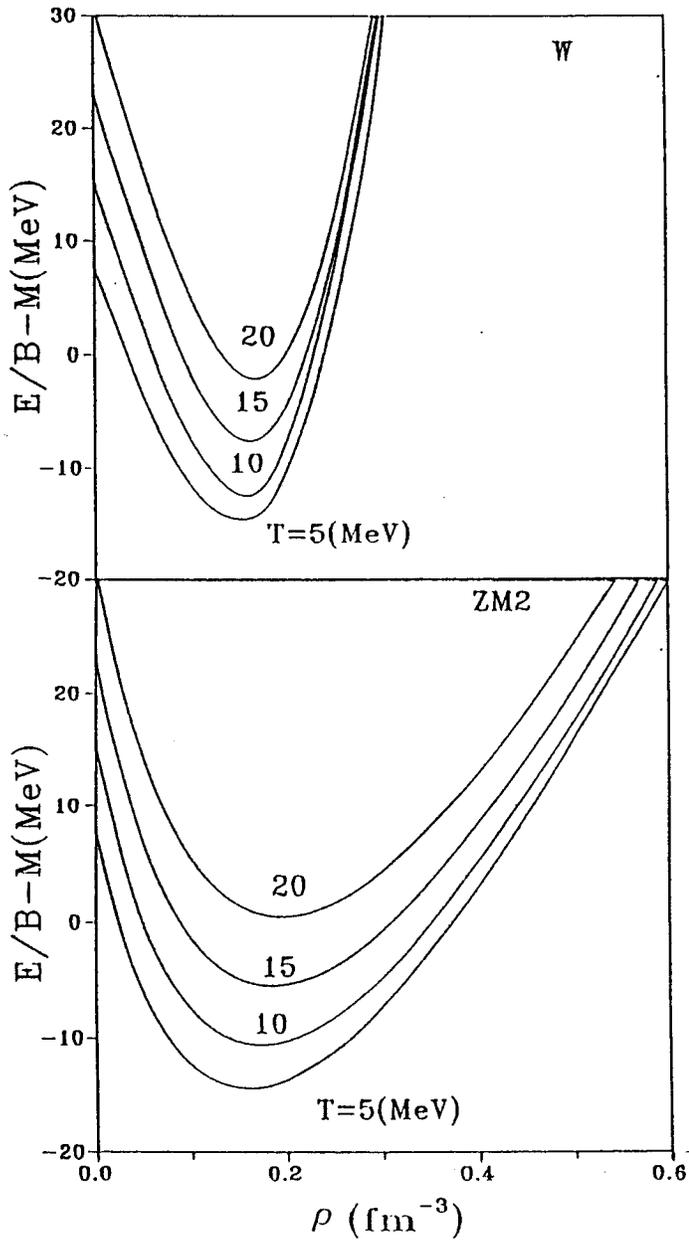
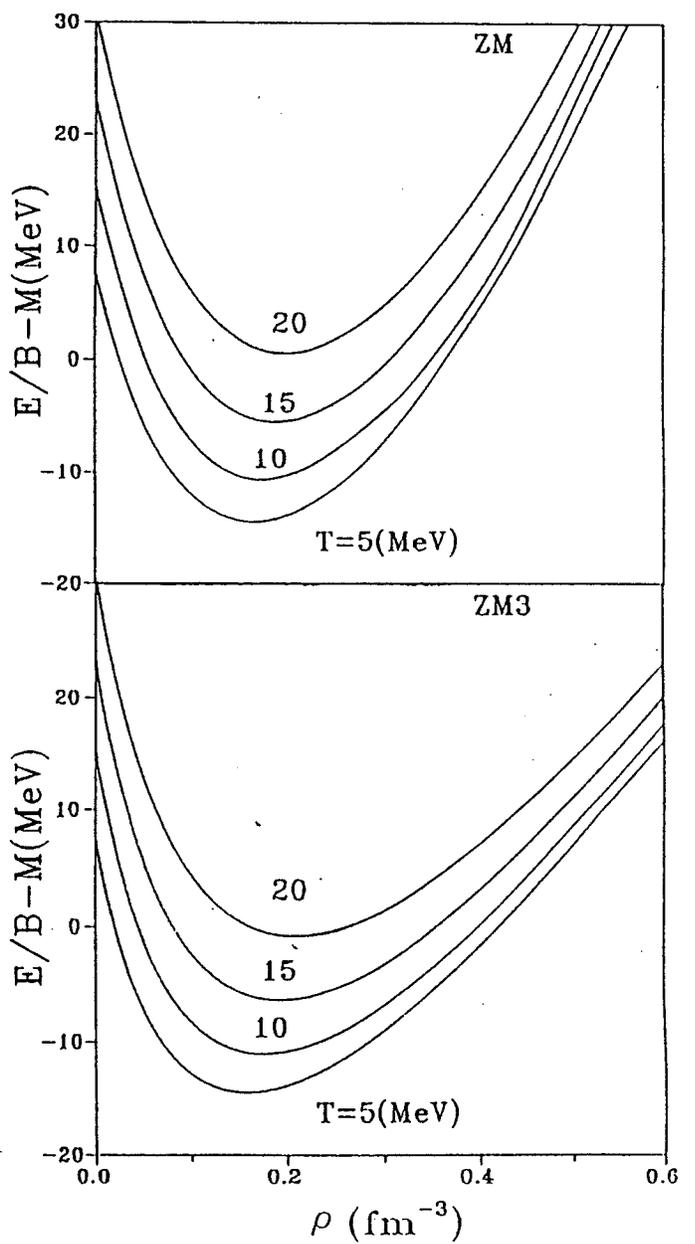
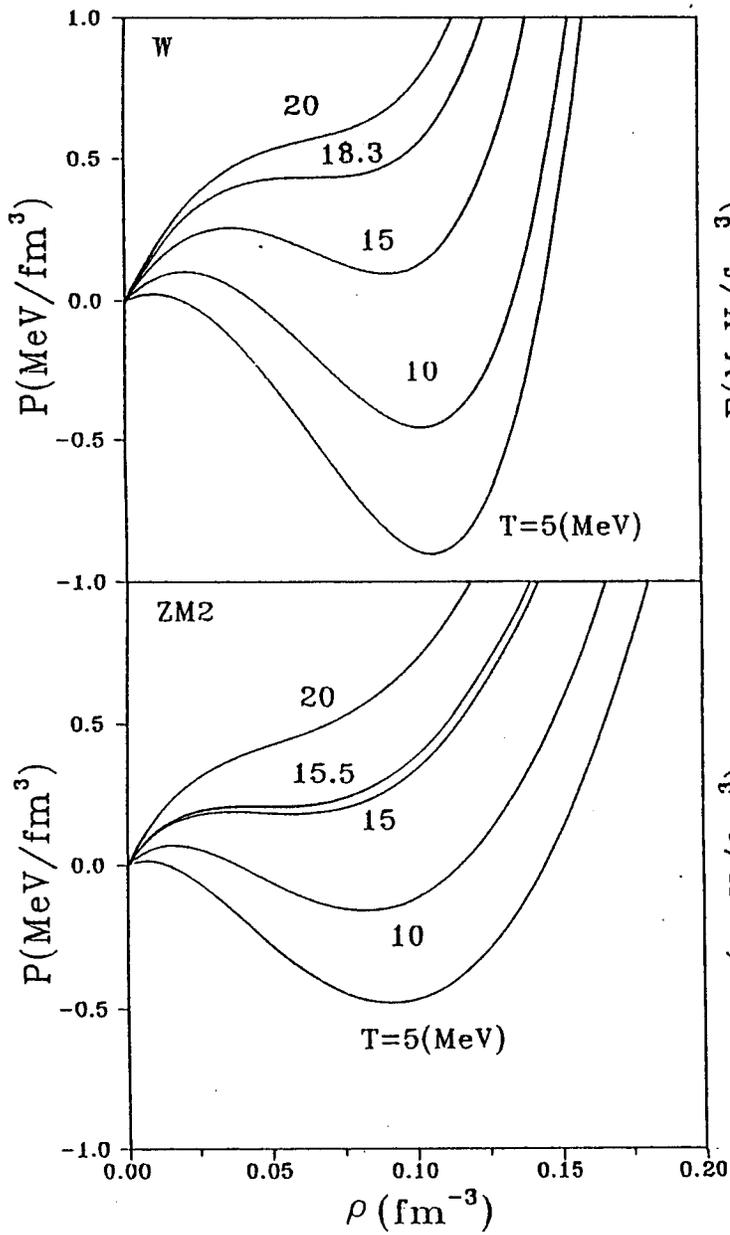
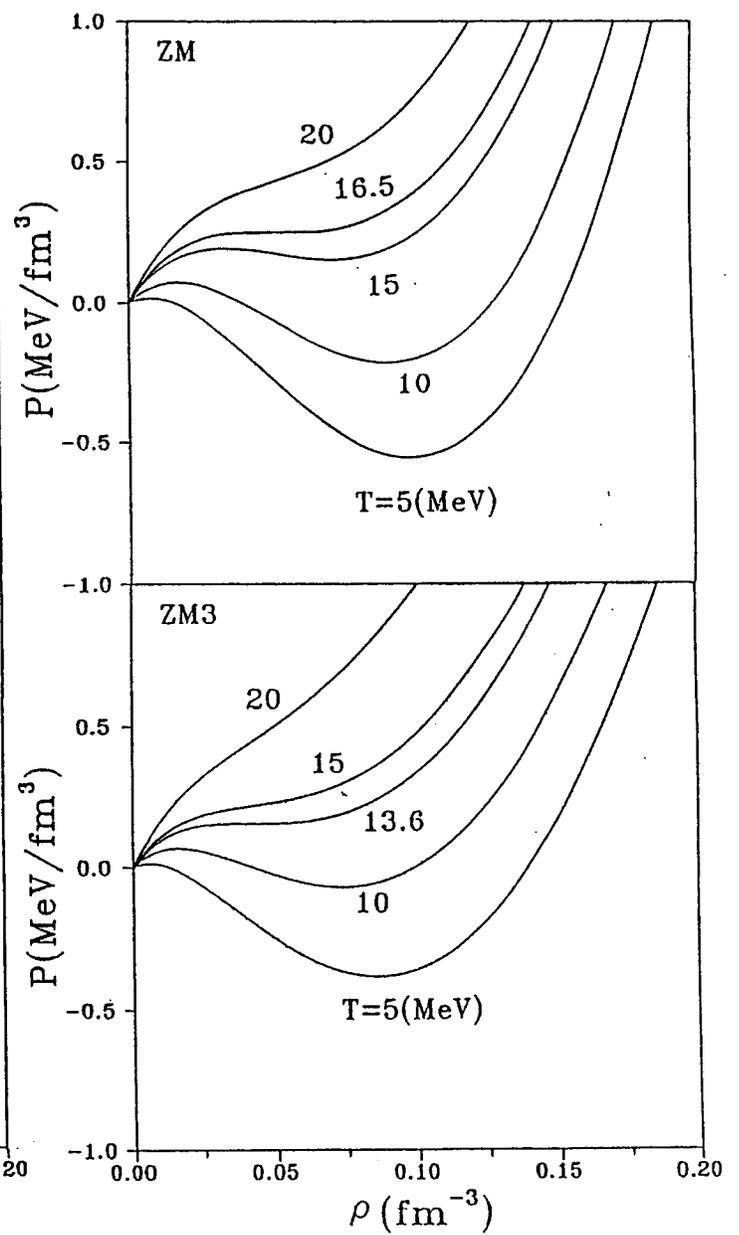
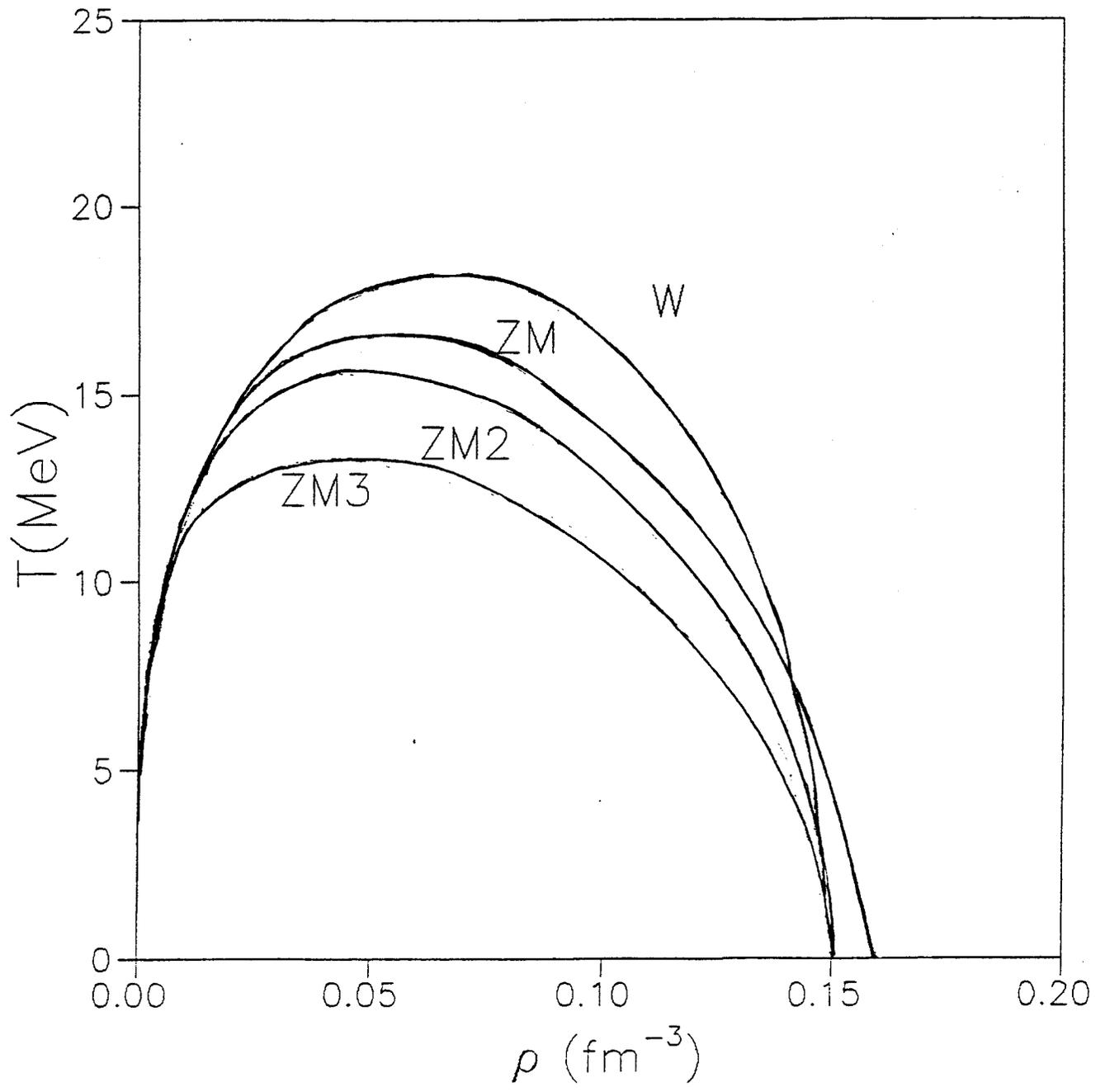
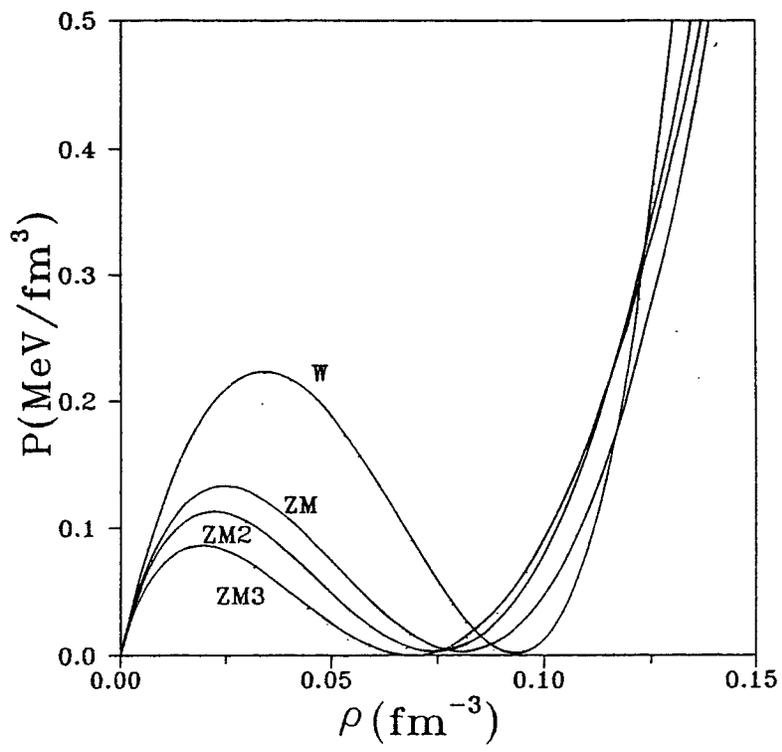
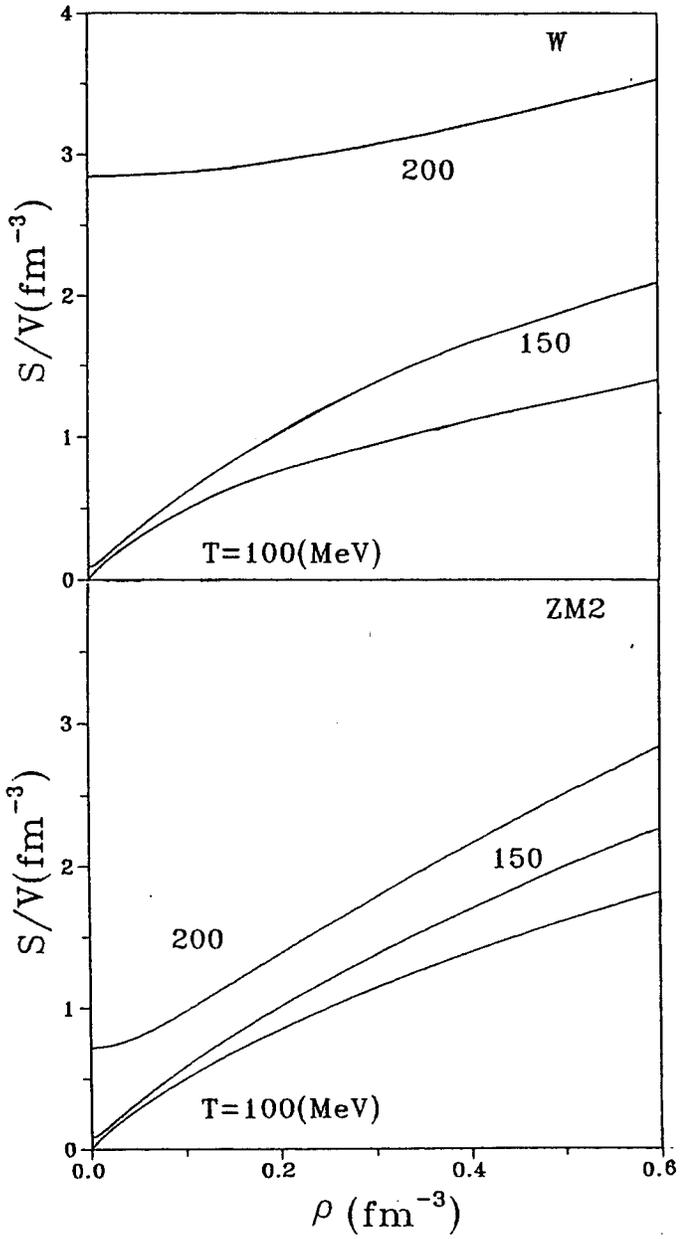
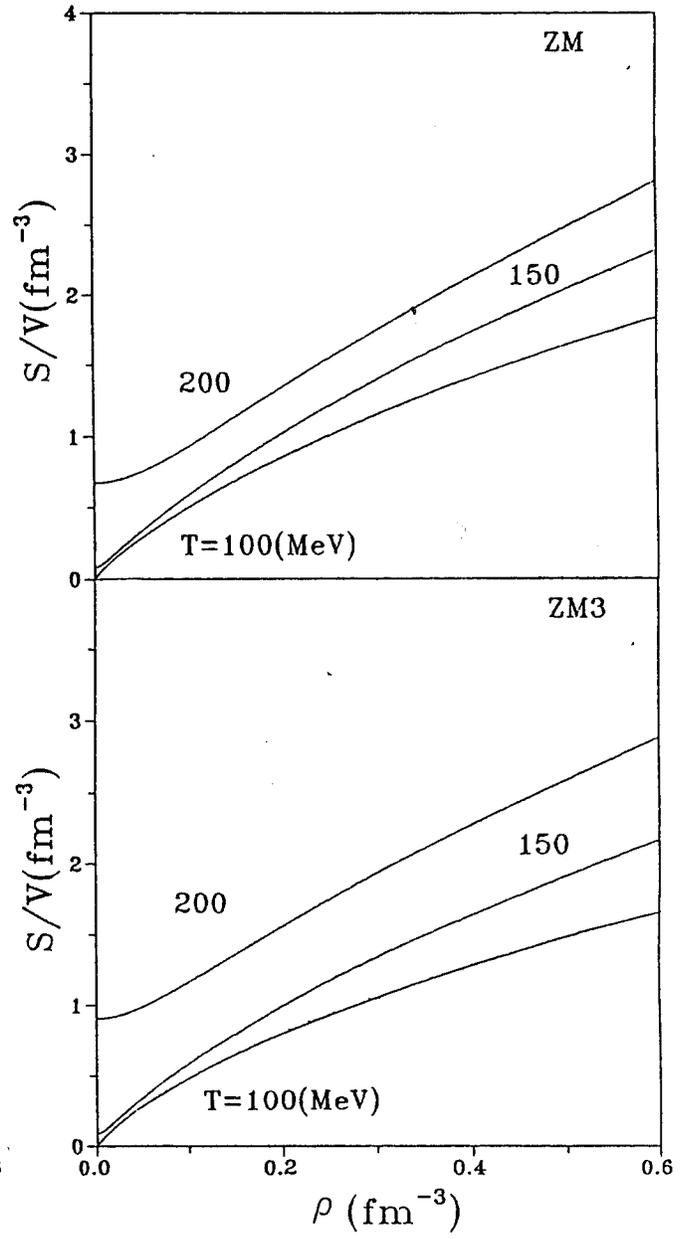
\end{document}